\documentclass[10pt]{article}
\usepackage{a4wide}
\usepackage{setspace}
\makeatletter

\newcommand{\LyX}{L\kern-.1667em\lower.25em\hbox{Y}\kern-.125emX\@}

\usepackage{latexsym}
\usepackage{amssymb}
\usepackage{amsfonts}
\makeatother

\begin{document}

{\par\centering \textbf{\large A Classification of Incomparable States}\\
\large \par}

{\par\centering Somshubhro Bandyopadhyay\( ^{\natural \star } \)\footnote{
som@ee.ucla.edu, dhom@boseinst.ernet.in
}, Vwani P. Roychowdhury\( ^{\natural } \)\footnote{
vwani@ee.ucla.edu
} and Ujjwal Sen\( ^{\star } \)\footnote{
dhom@boseinst.ernet.in
}\par}

{\par\centering {\small \( ^{\natural } \)}\emph{\small Electrical Engineering
Department, UCLA, Los Angeles, CA 90095, USA} {\small }\small \par}

{\par\centering \emph{\small \( ^{\star } \)Department of Physics, Bose Institute,
93/1 A. P. C Road, Kolkata-700009, INDIA} {\small }\small \par}

\begin{abstract}
Let \( \left\{ \left| \psi \right\rangle ,\left| \phi \right\rangle \right\}  \)
be an incomparable pair of states (\( \left| \psi \right\rangle \nleftrightarrow \left| \phi \right\rangle  \)),
\emph{i.e.}, \( \left| \psi \right\rangle  \) and \( \left| \phi \right\rangle  \)
cannot be transformed to each other with probability one by local transformations
and classical communication (LOCC). We show that incomparable states can be
multiple-copy transformable, \emph{i.e.}, there can exist a \emph{k}, such that
\( \left| \psi \right\rangle ^{\otimes k+1}\rightarrow \left| \phi \right\rangle ^{\otimes k+1} \),
i.e., \( k+1 \) copies of \( \left| \psi \right\rangle  \) can be transformed
to \( k+1 \) copies of \( \left| \phi \right\rangle  \) with probability one
by LOCC but \( \left| \psi \right\rangle ^{\otimes n}\nleftrightarrow \left| \phi \right\rangle ^{\otimes n}\; \forall n\leq k \).
We call such states \emph{k}-copy LOCC incomparable. We provide a necessary
condition for a given pair of states to be \emph{k}-copy LOCC incomparable for
some \emph{k}. We also show that there exist states that are neither \emph{k}-copy
LOCC incomparable for any \emph{k} nor catalyzable even with multiple copies.
We call such states strongly incomparable. We give a sufficient condition for
strong incomparability. 

We demonstrate that the optimal probability of a conclusive transformation involving
many copies, \( p_{max}\left( \left| \psi \right\rangle ^{\otimes m}\rightarrow \left| \phi \right\rangle ^{\otimes m}\right)  \)
can decrease exponentially with the number of source states \( m \), even if
the source state has \emph{more} entropy of entanglement. We also show that
the probability of a conclusive conversion might not be a monotonic function
of the number of copies. 
\end{abstract}
Fascinating developments in quantum information theory \cite{bennett} and quantum
computing \cite{preskill} during the past decade has led us to view entanglement
as a valued physical resource. Consequently, recent studies have largely been
devoted towards its quantification in appropriate limits (finite or asymptotic),
optimal manipulation, and transformation properties under local operations and
classical communication (LOCC) \cite{nielsen, vidal, vjn, JP, bennett2, lp}.
Since the specific tasks that can be accomplished with entanglement as a resource
is closely related to its transformation properties, it is of importance to
know what transformations are allowed under LOCC. Suppose Alice and Bob share
a pure state \( \left| \psi \right\rangle  \) (source state), which they wish
to convert to another entangled state \( \left| \phi \right\rangle  \) (target
state) under LOCC. A necessary and sufficient condition for this transformation
to be possible with certainty (denoted by \( \left| \psi \right\rangle \rightarrow \left| \phi \right\rangle  \))
has been obtained by Nielsen \cite{nielsen}. If such a deterministic transformation
is not possible but \( \left| \psi \right\rangle  \) has at least as many Schmidt
coefficients as \( \left| \phi \right\rangle  \), then one can still obtain
the target state with some non-zero probability of success. Such transformations
are said to be conclusive and an optimal local conversion strategy has been
obtained by Vidal \cite{vidal}. It should be noted that these transformations
are in the finite-copy regime and exact as well, in the sense, that the final
state is not an approximation of the desired target state. Recently deterministic
but approximate transformations in the finite-copy regime has also been studied
\cite{vjn}. 

A consequence of Nielsen's result is the existence of ``incomparable states''
\cite{nielsen}. A given pair \( \left\{ \left| \psi \right\rangle ,\left| \phi \right\rangle \right\}  \)
is said to be incomparable if none of them can be converted into the other with
probability one (denoted by \( \left| \psi \right\rangle \nleftrightarrow \left| \phi \right\rangle  \)).
However, Jonathan and Plenio \cite{JP} showed that for certain incomparable
pairs one can nevertheless perform entanglement assisted local operations and
classical communication (ELOCC), where transformation among previously incomparable
states is made possible (only one way, say \( \left| \psi \right\rangle \rightarrow \left| \phi \right\rangle  \))
using an auxiliary entangled state. The borrowed entanglement remains intact
in the process and can thus be used for further manipulations. This phenomenon
is what they termed ``catalysis'' and the auxiliary state, ``catalyst''.
This is an interesting twist to the problem of incomparable states because strictly
speaking, if \( \left\{ \left| \psi \right\rangle ,\left| \phi \right\rangle \right\}  \)
is a catalyzable pair, then their entanglement is no more incomparable, in the
sense that we can now say that \( \left| \psi \right\rangle  \) contains at
least as much entanglement as \( \left| \phi \right\rangle  \). 

The present work aims towards a better understanding of incomparable states
from their transformation properties under LOCC , strictly in the \emph{finite-copy}
regime. We ask: 

\emph{Does a given incomparable pair remain incomparable if multiple copies
of the source state take part in the transformation, to obtain} as many copies
\emph{of the target state under LOCC ?}

\emph{i.e.,} \emph{if \( \left| \psi \right\rangle \nleftrightarrow \left| \phi \right\rangle  \),
then can the transformation \( \left| \psi \right\rangle ^{\otimes k}\rightarrow \left| \phi \right\rangle ^{\otimes k} \)
(or \( \left| \phi \right\rangle ^{\otimes k}\rightarrow \left| \psi \right\rangle ^{\otimes k} \))
be realized with probability one by LOCC for some \( k>1 \) ?} 

It turns out that such states indeed exist. However there also exist incomparable
pairs that remain incomparable with \emph{k} copies for all \emph{k} even under
ELOCC. We thus obtain essentially new types of incomparable states and provide
a classification from their transformation properties. The main results are
the following: 

1. We show that there exist incomparable pairs \( \left\{ \left| \psi \right\rangle ,\left| \phi \right\rangle \right\}  \)
such that \( \left| \psi \right\rangle ^{\otimes (k+1)}\rightarrow \left| \phi \right\rangle ^{\otimes (k+1)} \)
(\( \rightarrow  \) read ``transforms to'') by LOCC for some finite \( k \)
with probability one but remain incomparable till \emph{k} copies i.e., \( \left| \psi \right\rangle ^{\otimes n}\nleftrightarrow \left| \phi \right\rangle ^{\otimes n}\; \forall n\leq k \).
We call such states \emph{k-copy} \emph{LOCC incomparable.} We also provide
a \emph{necessary} condition for \emph{k}-copy LOCC incomparability. Here one
may note that the existence of the transformation \( \left| \psi \right\rangle ^{\otimes (k+1)}\rightarrow \left| \phi \right\rangle ^{\otimes (k+1)} \)
\emph{}with certainty for some \emph{k} rules out the reverse possibility, \emph{i.e.},
\emph{\( \left| \phi \right\rangle ^{\otimes n}\nrightarrow \left| \psi \right\rangle ^{\otimes n}\; \forall n \). }

2. We show that there exist incomparable pairs \( \left\{ \left| \psi \right\rangle ,\left| \phi \right\rangle \right\}  \)
such that \( \left\{ \left| \psi \right\rangle ^{\otimes k},\left| \phi \right\rangle ^{\otimes k}\right\}  \)
remain incomparable for any \( k \) even under ELOCC. We call such states \emph{strongly
incomparable.} Obviously such states are also \emph{k}-copy LOCC incomparable
for any \emph{k}. We go on to provide a \emph{sufficient} condition for a pair
to be strongly incomparable in \( d\times d \) for all \( d\geq 3 \). This
gives an easy method of \emph{generating} such states in \( d\times d \). 

3. For certain incomparable pairs, increasing the number of copies results in
an \emph{exponential decrease} in the conclusive transformation probability
\cite{vidal}, i.e., \( p_{max}\left( \left| \psi \right\rangle ^{\otimes k}\rightarrow \left| \phi \right\rangle ^{\otimes k}\right) \sim c^{k},\; c<1 \),
even though the source state \( \left| \psi \right\rangle  \) has \emph{more}
entropy of entanglement \cite{bennett2}. This surprisingly shows that collective
manipulations need not necessarily be advantageous in the case of conclusive
transformations \cite{vidal, vjn}. 

4. We also demonstrate that the optimal probability of a conclusive conversion
\cite{vidal} need \emph{not} be a monotonic function of the number of copies,
i.e., \( p_{max}\left( \left| \psi \right\rangle ^{\otimes k}\rightarrow \left| \phi \right\rangle ^{\otimes k}\right)  \)
may not behave monotonically with \emph{}the number of copies \emph{k}. 

Let \( \left| \psi \right\rangle _{AB}=\sum ^{d}_{i=1}\sqrt{\alpha _{i}}\left| i\right\rangle _{A}\left| i\right\rangle _{B} \)
and \( \left| \phi \right\rangle _{AB}=\sum ^{d}_{i=1}\sqrt{\beta _{i}}\left| i\right\rangle _{A}\left| i\right\rangle _{B} \)
be bipartite pure states with Schmidt coefficients, \( \alpha _{1}\geq \alpha _{2}\geq ...\geq \alpha _{d}\geq 0 \)
and \( \beta _{1}\geq \beta _{2}\geq ...\geq \beta _{d}\geq 0 \), respectively.
The eigenvalues of the reduced density matrices \( \rho _{\psi }\equiv Tr_{B(or\, A)}\left( \left| \psi \right\rangle _{AB}\left\langle \psi \right| \right)  \)
and \( \rho _{\phi }\equiv Tr_{B(or\, A)}\left( \left| \phi \right\rangle _{AB}\left\langle \phi \right| \right)  \)
are \( \alpha _{1},\alpha _{2},...,\alpha _{d} \) and \( \beta _{1},\beta _{2},...,\beta _{d} \)
respectively. Denote the vector of the eigenvalues as \( \lambda _{\psi }\equiv \left( \alpha _{1},...,\alpha _{d}\right)  \)
and \( \lambda _{\phi }\equiv \left( \beta _{1},\beta _{2},...,\beta _{d}\right)  \).
We then have the following theorem due to Nielsen. 

\textbf{Theorem 1} \cite{nielsen}: \emph{\( \left| \psi \right\rangle  \)
transforms to \( \left| \phi \right\rangle  \) using LOCC with probability
one if and only if \( \lambda _{\psi } \) is majorized by \( \lambda _{\phi } \)
(written} \( \lambda _{\psi }\prec \lambda _{\phi } \)\emph{)}, \emph{that
is, if and only if} for each \( m \) in the range \( 1,...,d \), 
\begin{equation}
\label{1}
\sum ^{m}_{i=1}\lambda ^{(i)}_{\psi }\leq \sum ^{m}_{i=1}\lambda ^{(i)}_{\phi }
\end{equation}
If Nielsen's criterion is violated then a deterministic transformation is not
possible. However for such cases if the source state has at least as many non-zero
Schmidt coefficients as the target state, then a conclusive transformation is
possible with the optimal probability given by \( p_{max}\left( \left| \psi \right\rangle \rightarrow \left| \phi \right\rangle \right) =\min _{1\leq l\leq d}\frac{E_{l}\left( \left| \psi \right\rangle \right) }{E_{l}\left( \left| \phi \right\rangle \right) } \),
where \( E_{l}\left( \left| \psi \right\rangle \right) =1-\sum ^{l-1}_{i=1}\alpha _{i} \)
\cite{vidal}. 

A consequence of Nielsen's result is the existence of incomparable states \cite{nielsen};
the states that are not transformable to one another with 100\% probability
by LOCC as shown in the following example:
\begin{eqnarray}
\left| \psi \right\rangle  & = & \sqrt{0.4}\left| 00\right\rangle +\sqrt{0.36}\left| 11\right\rangle +\sqrt{0.14}\left| 22\right\rangle +\sqrt{0.1}\left| 33\right\rangle \label{2} \\
\left| \phi \right\rangle  & = & \sqrt{0.5}\left| 00\right\rangle +\sqrt{0.25}\left| 11\right\rangle +\sqrt{0.25}\left| 22\right\rangle \label{3} 
\end{eqnarray}
 From Nielsen's theorem it follows that these two states are incomparable, \emph{i.e.},
neither \( \left| \psi \right\rangle \rightarrow \left| \phi \right\rangle  \)
nor \( \left| \phi \right\rangle \rightarrow \left| \psi \right\rangle  \)
by LOCC. 

Let \( \left\{ \left| \psi \right\rangle ,\left| \phi \right\rangle \right\}  \)
be an incomparable pair. Suppose now we have \emph{k} copies of the source state
at our disposal. Is it possible to obtain \emph{k} copies of the target state
after some collective manipulation, for some finite \emph{k} by LOCC ? The answer
to this question is both yes and no. We first consider the ``yes'' case and
illustrate with a simple example. We take up the ``no'' case later. 

Consider the incomparable pair given in Eqs. (2) and (3). We now show that two
copies of \( \left| \psi \right\rangle  \) can be transformed to two copies
of \( \left| \phi \right\rangle  \) by LOCC, i.e., the transformation \( \left| \psi \right\rangle ^{\otimes 2}\rightarrow \left| \phi \right\rangle ^{\otimes 2} \)
is indeed possible by LOCC with probability one. The corresponding \( \lambda  \)-vectors
are

\begin{eqnarray}
\lambda _{\psi ^{\otimes 2}} & = & \left( .16,.144,.144,.1296,.056,.056,.0504,.0504,.04,.04,.036,.036,.0196,.014,.014,.001\right) \quad \; \label{4} \\
\lambda _{\phi ^{\otimes 2}} & = & \left( .25,.125,.125,.125,.125,.0625,.0625,.0625,.0625,0,0,0,0,0,0,0\right) \label{5} 
\end{eqnarray}
 It is easy to check that \( \lambda _{\psi ^{\otimes 2}} \) \( \prec  \)
\( \lambda _{\phi ^{\otimes 2}} \) implying that the transformation \( \left| \psi \right\rangle ^{\otimes 2}\rightarrow \left| \phi \right\rangle ^{\otimes 2} \)
can in fact be realized by LOCC with certainty. 

\textbf{Definition 1:} \emph{An incomparable pair, say \( \left\{ \left| \psi \right\rangle ,\left| \phi \right\rangle \right\}  \)
is said to be} \emph{k-copy LOCC incomparable if either \( \left| \psi \right\rangle ^{\otimes \left( k+1\right) }\rightarrow \left| \phi \right\rangle ^{\otimes \left( k+1\right) }\: or\: \left| \phi \right\rangle ^{\otimes \left( k+1\right) }\rightarrow \left| \psi \right\rangle ^{\otimes \left( k+1\right) } \)
under LOCC and \( \left| \psi \right\rangle ^{\otimes n}\nleftrightarrow \left| \phi \right\rangle ^{\otimes n}\; \forall n\leq k \).}

Note that there cannot exist any incomparable pair \emph{\( \left\{ \left| \psi \right\rangle ,\left| \phi \right\rangle \right\}  \)}
for which \emph{\( \left| \psi \right\rangle ^{\otimes \left( k+1\right) }\rightarrow \left| \phi \right\rangle ^{\otimes \left( k+1\right) } \)}
and \( \left| \phi \right\rangle ^{\otimes \left( k+1\right) }\rightarrow \left| \psi \right\rangle ^{\otimes \left( k+1\right) } \)
hold simultaneously. This follows from the fact that if both transformations
hold, then the Schmidt coefficients of the states \emph{\( \left| \psi \right\rangle  \)}
and \emph{\( \left| \phi \right\rangle  \)} must be equal, which is a contradiction. 

Examples of \emph{k-copy LOCC incomparable} pairs: The first example that we
have given is single-copy LOCC incomparable. We now provide examples for values
of \emph{k} not equal to one. 

(a) \emph{2-copy LOCC incomparable pair:} 

\begin{eqnarray}
\left| \psi \right\rangle  & = & \sqrt{0.4}\left| 00\right\rangle +\sqrt{0.4}\left| 11\right\rangle +\sqrt{0.1}\left| 22\right\rangle +\sqrt{0.1}\left| 33\right\rangle \label{6} \\
\left| \phi \right\rangle  & = & \sqrt{0.5}\left| 00\right\rangle +\sqrt{0.27}\left| 11\right\rangle +\sqrt{0.23}\left| 22\right\rangle \label{7} 
\end{eqnarray}
 Observe that \( p_{max}\left( \left| \psi \right\rangle \rightarrow \left| \phi \right\rangle \right) \cong 87\% \)
and \( p_{max}\left( \left| \psi \right\rangle ^{\otimes 2}\rightarrow \left| \phi \right\rangle ^{\otimes 2}\right) \cong 99\% \).
Therefore, as we might expect, transformation probability increases with number
of copies. One can easily check, the probability becomes one with three copies
of the source state \emph{i.e.}, \( \lambda _{\psi ^{\otimes 3}}\prec \lambda _{\phi ^{\otimes 3}} \)
and Nielsen's theorem then implies \( \left| \psi \right\rangle ^{\otimes 3}\rightarrow \left| \phi \right\rangle ^{\otimes 3} \)
by LOCC with probability one. 

(b) \emph{5-copy LOCC incomparable pair:}
\begin{eqnarray}
\left| \psi \right\rangle  & = & \sqrt{0.4}\left| 00\right\rangle +\sqrt{0.4}\left| 11\right\rangle +\sqrt{0.1}\left| 22\right\rangle +\sqrt{0.1}\left| 33\right\rangle \label{8} \\
\left| \phi \right\rangle  & = & \sqrt{0.48}\left| 00\right\rangle +\sqrt{0.27}\left| 11\right\rangle +\sqrt{0.25}\left| 22\right\rangle \label{9} 
\end{eqnarray}

Existence of such transformations might be useful sometimes. Take for instance
the incomparable pair given in Eqs. (2) and (3), which is \emph{single-copy
LOCC incomparable}. This pair is catalyzable as well (one can find without much
difficulty an appropriate catalyst for this pair, for example the state \( \left| \chi \right\rangle =0.6\left| 44\right\rangle +0.4\left| 55\right\rangle  \)
is a valid catalyst for the pair). Thus to obtain \emph{two} copies of \( \left| \phi \right\rangle  \)
from \emph{two} copies of \( \left| \psi \right\rangle  \) one needs \emph{two}
such entanglement assisted transformations. But we have shown that the same
goal can be reached by a single collective transformation without any catalyst. 

It may be noted that in all the above examples, \( p_{max}\left( \left| \psi \right\rangle ^{\otimes n}\rightarrow \left| \phi \right\rangle ^{\otimes n}\right) \geq p_{max}\left( \left| \psi \right\rangle ^{m}\rightarrow \left| \phi \right\rangle ^{m}\right)  \)
\( \forall n,m;\; m\leq n\leq k+1 \) for some \( k \). Thus we have in these
examples, as one would also expect, an increase in transformation probability
with the number of copies which becomes unity for \( n=k+1 \) for some \( k \).
This is intuitively satisfying because while increasing the number of copies
we are accumulating ``more'' entanglement on the source side. It is worth
stressing that this monotonicity might only be expected when one can \emph{a
priori} say \emph{definitively} that the pair is \emph{k-copy LOCC incomparable},
for some \emph{k}. We return to this point later in this paper. 

Let \( \left\{ \left| \psi \right\rangle ,\left| \phi \right\rangle \right\}  \)
be \emph{}a \emph{k}-copy LOCC incomparable \emph{}pair. Then \emph{\( \left| \psi \right\rangle ^{\otimes (k+1)}\rightarrow \left| \phi \right\rangle ^{\otimes (k+1)} \)
,} i.e., \emph{k}+1 copies of \( \left| \psi \right\rangle  \) transforms to
\emph{k}+1 \emph{}copies of \( \left| \phi \right\rangle  \), by LOCC. Can
we say that \emph{n} copies of \( \left| \psi \right\rangle  \) transforms
to \emph{n} copies of \( \left| \phi \right\rangle  \) when \emph{n} is greater
than \emph{k}+1? 

\textbf{Conjecture:} If \( \left| \psi \right\rangle ^{\otimes k+1}\rightarrow \left| \phi \right\rangle ^{\otimes k+1} \)
by LOCC then \( \left| \psi \right\rangle ^{\otimes n}\rightarrow \left| \phi \right\rangle ^{\otimes n}\; \forall n\geq k+2 \). 

Of course \( \left| \psi \right\rangle ^{\otimes n}\rightarrow \left| \phi \right\rangle ^{\otimes n} \)
when \( n \) is an integral multiple of \( k+1 \). Hence the cases of interest
are those values of \( n\geq k+2,\textrm{ where }n\neq m\left( k+1\right) ,\, m \)
is an integer. 

We would now like to ask what condtions need to be satisfied for \emph{k}-copy
LOCC incomparability. Suppose \( \left\{ \left| \psi \right\rangle ,\left| \phi \right\rangle \right\}  \)
be an incomparable pair. Would a transformation \( \left| \psi \right\rangle ^{\otimes k}\rightarrow \left| \phi \right\rangle ^{\otimes k} \)
(or \emph{vice versa}) be always possible for some \( k \) by LOCC or even
by ELOCC ? We now give a \emph{necessary} condition for such transformations
to exist. 

\textbf{Lemma 1:} \emph{Let \( \left| \psi \right\rangle  \) and \( \left| \phi \right\rangle  \)
be \( d\times d \) states, with ordered Schmidt coefficients \( \left\{ \alpha _{j}\right\} ,\left\{ \beta _{j}\right\} ,1\leq j\leq d \)
respectively. Then there exists some \( k>1 \) such that \( \left| \psi \right\rangle ^{\otimes k}\rightarrow \left| \phi \right\rangle ^{\otimes k} \)
under LOCC only if \( \alpha _{1}\leq \beta _{1} \) and \( \alpha _{d}\geq \beta _{d} \).
The same necessary condition also holds for \( \left| \psi \right\rangle ^{\otimes k}\rightarrow \left| \phi \right\rangle ^{\otimes k} \)
under ELOCC. }

Proof. If \emph{}there is some \emph{k} such that \( \left| \psi \right\rangle ^{\otimes k}\rightarrow \left| \phi \right\rangle ^{\otimes k} \),
then from Nielsen's theorem it follows that \( \alpha ^{k}_{1}\leq \beta ^{k}_{1} \)
and \( 1-\alpha ^{k}_{d}\leq 1-\beta ^{k}_{d} \) which implies \emph{\( \alpha _{1}\leq \beta _{1} \)}
and \( \alpha _{d}\geq \beta _{d} \). This proves the first part of the lemma. 

It has been shown in Ref. \cite{JP} that \( \left| \psi \right\rangle \rightarrow \left| \phi \right\rangle  \)
under ELOCC only if \emph{\( \alpha _{1}\leq \beta _{1} \)} and \( \alpha _{d}\geq \beta _{d} \).
It is straightforward to show that similar condition holds for \( \left| \psi \right\rangle ^{\otimes k}\rightarrow \left| \phi \right\rangle ^{\otimes k} \)
under ELOCC. \( \Box  \)

From lemma 1 it follows that \( 3\times 3 \) incomparable states remain incomparable
even if \emph{multiple} copies are available. This is due to the fact that for
\( 3\times 3 \) incomparable states if \emph{\( \alpha _{1}<\beta _{1}, \)}
then \( \alpha _{3}<\beta _{3} \) . Hence incomparable states in \( 3\times 3 \)
are neither catalyzable even with multiple copies nor multiple-copy transformable. 

The existence of incomparable states that are neither \emph{k-copy LOCC incomparable}
for any \emph{k} nor \emph{catalyzable} with multiple copies allows us to define
such states in a general way. 

\textbf{Definition 2:} \emph{An incomparable pair} \( \left\{ \left| \psi \right\rangle ,\left| \phi \right\rangle \right\}  \)
\emph{is said to be strongly incomparable if the pair is non-catalyzable even
with multiple copies, i.e., \( \left| \psi \right\rangle ^{\otimes k}\nleftrightarrow \left| \phi \right\rangle ^{\otimes k} \)}
\emph{under ELOCC for all k. }

Strongly incomparable pairs are obviously \emph{k}-copy LOCC incomparable for
all \emph{k}. The following result provides a sufficient condition for strong
incomparability. 

\textbf{Theorem 2:} \emph{Let \( \left| \psi \right\rangle  \) and \( \left| \phi \right\rangle  \)
be \( d\times d \) states, with ordered Schmidt coefficients \( \left\{ \alpha _{j}\right\} ,\left\{ \beta _{j}\right\} ,1\leq j\leq d \)
respectively. A sufficient condition that they form a strongly incomparable
pair is \( \alpha _{1}<\beta _{1} \) and} \( \alpha _{d}<\beta _{d} \) \emph{OR}
\emph{\( \alpha _{1}>\beta _{1} \) and} \( \alpha _{d}>\beta _{d} \). 

Proof. From lemma 1 it follows that if \( \alpha _{1}<\beta _{1} \) and \emph{\( \alpha _{d}<\beta _{d} \)}
OR \emph{}\( \alpha _{1}>\beta _{1} \) and \( \alpha _{d}>\beta _{d} \) then
\emph{\( \left| \psi \right\rangle ^{\otimes k}\nleftrightarrow \left| \phi \right\rangle ^{\otimes k} \)}
under ELOCC for all \( k \). Hence the proof. \( \Box  \)

As noted earlier, the incomparable states in \( 3\times 3 \) are strongly incomparable.
Theorem 2 provides a method for constructing strongly incomparable states in
\( d\times d \) for any \( d\geq 3 \). 

If \( \left\{ \left| \psi \right\rangle ,\left| \phi \right\rangle \right\}  \)
is strongly incomparable then there does not exist any local strategy such that
\emph{k} copies of \( \left| \psi \right\rangle  \) can be converted into \emph{k}
copies of \( \left| \phi \right\rangle  \) or \emph{vice versa} with certainty
under LOCC and even by ELOCC. Hence for any \emph{k,} the transformation \( \left| \psi \right\rangle ^{\otimes k}\rightarrow \left| \phi \right\rangle ^{\otimes k} \)
(or the reverse one) is necessarily conclusive. We would now like to know how
the transformation probability changes with the number of copies when the conversion
is conclusive. Before that we prove some general results. 

Let \( \left| \psi \right\rangle  \) and \( \left| \phi \right\rangle  \)
be \( d\times d \) states with ordered Schmidt coefficients \( \left\{ \alpha _{j}\right\} ,\left\{ \beta _{j}\right\} ,1\leq j\leq d \)
respectively and \( \left| MES\right\rangle  \) be a maximally entangled state
in \( d\times d \).  

\textbf{Lemma 2:} \emph{\( p_{max}\left( \left| \psi \right\rangle \rightarrow \left| MES\right\rangle \right) <p_{max}\left( \left| \phi \right\rangle \rightarrow \left| MES\right\rangle \right)  \)
iff} \( \alpha _{d}<\beta _{d} \). 

The proof follows by noting that \( p_{max}\left( \left| \psi \right\rangle \rightarrow \left| MES\right\rangle \right) =d\alpha _{d} \)
and \( p_{max}\left( \left| \phi \right\rangle \rightarrow \left| MES\right\rangle \right) =d\beta _{d} \)
\cite{vidal, lp}. We now show that the condition \emph{\( p_{max}\left( \left| \psi \right\rangle \rightarrow \left| MES\right\rangle \right) <p_{max}\left( \left| \phi \right\rangle \rightarrow \left| MES\right\rangle \right)  \)}
is sufficient to ensure that the optimal probability of a conclusive transformation
can never increase with the number of copies. 

\textbf{Theorem 3:} \emph{If \( p_{max}\left( \left| \psi \right\rangle \rightarrow \left| MES\right\rangle \right) <p_{max}\left( \left| \phi \right\rangle \rightarrow \left| MES\right\rangle \right)  \)
then \( p_{max}\left( \left| \psi \right\rangle ^{\otimes k}\rightarrow \left| \phi \right\rangle ^{\otimes k}\right)  \)
falls off exponentially with the number of copies. }

Proof. Let \( p_{max}\left( \left| \psi \right\rangle \rightarrow \left| MES\right\rangle \right) <p_{max}\left( \left| \phi \right\rangle \rightarrow \left| MES\right\rangle \right)  \).
Then from lemma 2 we have \( p_{max}\left( \left| \psi \right\rangle \rightarrow \left| \phi \right\rangle \right) \leq  \)\( \frac{\alpha _{d}}{\beta _{d}}<1 \).
Hence \( p_{max}\left( \left| \psi \right\rangle ^{\otimes k}\rightarrow \left| \phi \right\rangle ^{\otimes k}\right) =\min _{l\leq d^{k}}\frac{E_{l}\left( \left| \psi \right\rangle ^{\otimes k}\right) }{E_{l}\left( \left| \phi \right\rangle ^{\otimes k}\right) }\leq \frac{E_{d^{k}}\left( \left| \psi \right\rangle ^{\otimes k}\right) }{E_{d^{k}}\left( \left| \phi \right\rangle ^{\otimes k}\right) }=\left( \frac{\alpha _{d}}{\beta _{d}}\right) ^{k}<1,\forall k\geq 2 \).
\( \Box  \) 

One can show that under the condition of theorem 3 the optimal probability for
the transformation \emph{\( \left| \psi \right\rangle ^{\otimes k}\rightarrow \left| \phi \right\rangle ^{\otimes k} \)}
falls off exponentially with \emph{k} even under ELOCC. 

From lemma 2 it follows that if \emph{\( p_{max}\left( \left| \psi \right\rangle \rightarrow \left| \phi \right\rangle \right) \leq  \)\( \frac{\alpha _{d}}{\beta _{d}}<1 \)}
then \( p_{max}\left( \left| \psi \right\rangle \rightarrow \left| MES\right\rangle \right) <p_{max}\left( \left| \phi \right\rangle \rightarrow \left| MES\right\rangle \right)  \),
whereby theorem 3 implies that \( p_{max}\left( \left| \psi \right\rangle ^{\otimes k}\rightarrow \left| \phi \right\rangle ^{\otimes k}\right)  \)
falls off exponentially with the number of copies. \emph{}We stress that lemma
2 and theorem 3 hold irrespective of whether the states are incomparable or
not. From theorem 2 it follows that if \emph{\( \alpha _{1}<\beta _{1} \)}
and \emph{}\( \alpha _{d}<\beta _{d} \) then the states are strongly incomparable.
Let us also note that in Ref. \cite{JP} it was shown that if \emph{\( p_{max}\left( \left| \psi \right\rangle \rightarrow \left| \phi \right\rangle \right) = \)\( \frac{\alpha _{d}}{\beta _{d}} \),}
then probability of conclusive transformation cannot be increased in presence
of any catalyst. Thus we see that there exist  incomparable pairs for which
the conclusive transformation probability cannot be improved in presence of
any catalyst and using multiple copies the probability falls off exponentially.

Sometimes the condition \( p_{max}\left( \left| \psi \right\rangle \rightarrow \left| MES\right\rangle \right) <p_{max}\left( \left| \phi \right\rangle \rightarrow \left| MES\right\rangle \right)  \)
might be satisfied even though \( E\left( \left| \psi \right\rangle \right) >E\left( \left| \phi \right\rangle \right)  \)
where \emph{E} is the entropy of entanglement. Consider the following incomparable
pair in \( 3\times 3 \), which we know to be strongly incomparable:

\begin{eqnarray}
\left| \zeta \right\rangle  & = & \sqrt{0.4}\left| 00\right\rangle +\sqrt{0.4}\left| 11\right\rangle +\sqrt{0.2}\left| 22\right\rangle \label{12} \\
\left| \omega \right\rangle  & = & \sqrt{0.5}\left| 00\right\rangle +\sqrt{0.25}\left| 11\right\rangle +\sqrt{0.25}\left| 22\right\rangle \label{13} 
\end{eqnarray}
 We are interested in how the conclusive transformation probability \( p_{max}\left( \left| source\right\rangle ^{\otimes k}\rightarrow \left| target\right\rangle ^{\otimes k}\right)  \)
scales with \emph{k, k} being \emph{}however large but \emph{finite}. Let us
first collect the following facts about the above pair: 

1. \( E\left( \left| \zeta \right\rangle \right) >E\left( \left| \omega \right\rangle \right)  \),
which means that in the asymptotic limit, \( \left| \zeta \right\rangle  \)
generates a larger number of maximally entangled states as compared to \( \left| \omega \right\rangle  \)
\cite{bennett2}. 

2. Let \( \left| MES\right\rangle  \) be a maximally entangled state in \( 3\times 3 \).
\emph{}Then \emph{\( p_{max}\left( \left| \zeta \right\rangle \rightarrow \left| MES\right\rangle \right) <p_{max}\left( \left| \omega \right\rangle \rightarrow \left| MES\right\rangle \right)  \),}
which means that given a \emph{large but finite} number of copies one can obtain
more maximally entangled states from \( \left| \omega \right\rangle  \) when
we use a conclusive conversion protocol. 

\emph{Case 1.} \emph{\( \left| \zeta \right\rangle ,\left| \omega \right\rangle  \)
being the source and target states respectively: }

First note that \( p_{max}\left( \left| \zeta \right\rangle \rightarrow \left| \omega \right\rangle \right) =\frac{\alpha _{3}}{\beta _{3}}=\frac{4}{5} \).
Hence \( p_{max}\left( \left| \zeta \right\rangle ^{\otimes k}\rightarrow \left| \omega \right\rangle ^{\otimes k}\right) \leq \left( \frac{\alpha _{3}}{\beta _{3}}\right) ^{k}=\left( \frac{4}{5}\right) ^{k} \).
Therefore for large \( k \), \( p_{max}\left( \left| \zeta \right\rangle ^{\otimes k}\rightarrow \left| \omega \right\rangle ^{\otimes k}\right)  \)
falls off exponentially to zero even though \( E\left( \left| \zeta \right\rangle \right) >E\left( \left| \omega \right\rangle \right)  \).
Since the conversion is \emph{conclusive}, a successful conversion always results
in an \emph{exact} outcome. At this point it is instructive to analyze this
result by comparing it to an asymptotic conversion. Note that there is no contradiction
with the result of Bennett \emph{et al.} \cite{bennett2}. To see this, consider
what happens in an asymptotic conversion. It was shown in Ref. \cite{bennett2}
that in an asymptotic conversion, the yield approaches \( \frac{E\left( \xi \right) }{E\left( \omega \right) } \),
the fidelity approaching 1 and the success probability also approaching 1 in
the limit of large \emph{k}. Since \( \frac{E\left( \xi \right) }{E\left( \omega \right) }>1 \),
in the large \emph{k} limit we would obtain at least \emph{k} copies of \emph{\( \left| \omega \right\rangle  \).}
This apparent contradiction is resolved at once by noting that for any finite
\emph{k} however large, the conversion is always \emph{approximate} and the
success probability is always \emph{less} than 1. 

\emph{Case 2}. \emph{\( \left| \omega \right\rangle ,\left| \zeta \right\rangle  \)
being the source and target states respectively:}

We present this case through numerical results that indicate a rather surprising
feature. We find that as we keep increasing the number of copies the transformation
probability shows an approximately damped oscillatory behavior (see Fig. 1).
This clearly shows that the transformation probability may not be a monotonic
function of the number of copies. Note that the maximum transformation probability
occurs when \emph{k}=3. So the transformation probability increases to maximum
at \emph{k}=3 and then decays in an oscillatory fashion. What is curious in
this behaviour is the lack of monotonicity. Observe that with two copies, the
probability is less than that of three copies which in turn is greater than
that with four copies. 

To summarize, we have obtained new types of incomparable entanglement and proposed
a possible classification of incomparable states. In the first class we have
those incomparable pairs that admit deterministic transformation (under LOCC)
between themselves (one way) by using multiple copies. The second class consists
of those states that do not allow deterministic transformation between them
even when multiple copies are provided and catalysts are allowed. We call such
states strongly incomparable. Exact transformation between these states therefore
seems to be inherently probabilistic. For such transformations we have obtained
the following in the multiple-copy scenario: (1) collective operations need
not always be advantageous for conclusive transformations even when the source
state has more entropy of entanglement; (2) conclusive transformation probability
might not be a monotonic function of the number of copies. 

Our results open up many interesting open questions. A \emph{k}-copy LOCC incomparable
pair is (\emph{k}+1)-copy transformable. Based on numerical evidences we have
conjectured that it is also (\emph{k}+\emph{m})-copy transformable for all \emph{m}.
But an analytical proof is still wanting. It would also be desirable to know
whether the sufficient condition of strong incomparability is also a necessary
condition. \\

We would like to thank Tal Mor and Guruprasad Kar for useful discussions. The
work of S.B. and V.R. was sponsored in part by the Defense Advanced Research
Projects Agency (DARPA) project MDA 972-99-1-0017 (note that the content of
this paper does not necessarily reflect the position or the policy of the government
and no official endorsement should be inferred), and in part by the U.S. Army
Research Office/DARPA under contract/grant number DAAD 19-00-1-0172. U.S. acknowledges
partial support by the Council of Scientific and Industrial Research, Government
of India, New Delhi.

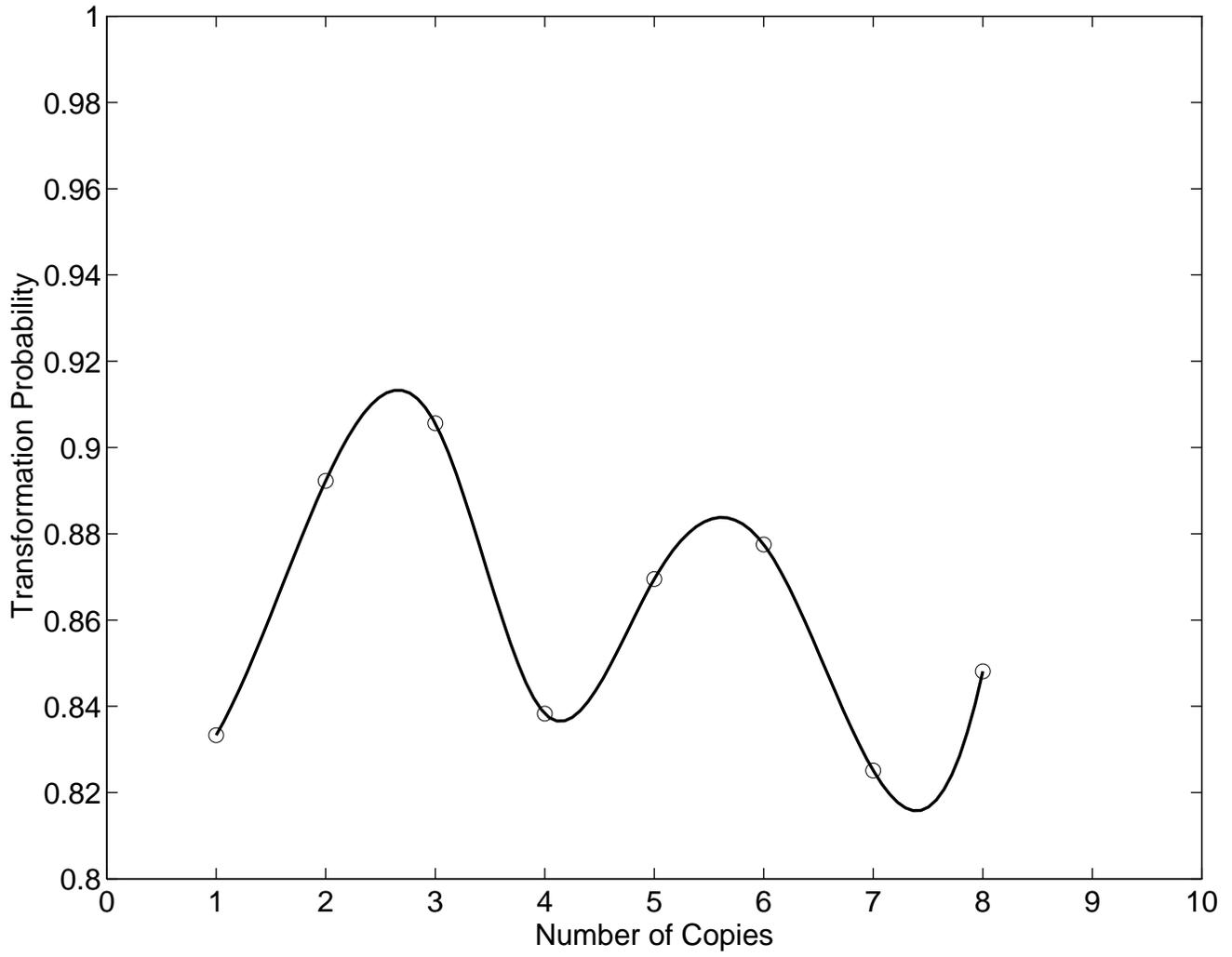
\end{document}